\documentclass[11pt]{article}

\usepackage[margin=1in]{geometry}
\usepackage{amsmath,amssymb}
\usepackage{graphicx}
\usepackage[colorlinks=true,citecolor=blue,linkcolor=blue,urlcolor=blue]{hyperref}
\usepackage[round]{natbib}

\title{\textbf{Grover Search for a Threshold-Fraction Variant of the Travelling Salesman Problem: A Circuit-Level Construction}}
\author{Aayra Dhindsa Sawkmie}
\date{}

\begin{document}
\maketitle

\begin{abstract}
The Travelling Salesman Problem (TSP) is a classic combinatorial optimisation problem which is NP-hard in optimisation form and NP-complete as a decision problem. The standard question asks for the shortest tour that visits every city exactly once and returns to the origin. This paper studies a threshold-fraction variant, which we call the robust TSP, formulated as a promise problem: given a distance threshold $K$ and a promised fraction $p$, decide whether at least a fraction $p$ of all tours have total length at most $K$, or whether no such tour exists. We show that Grover's quantum search algorithm decides this problem using $O(1/\sqrt{p})$ oracle queries, independent of the number of cities $N$, with each query implemented by a quantum circuit of polynomial size in $N$. For any fixed robustness level $p$, the total runtime is therefore polynomial in $N$. Measured in oracle queries, the quantum algorithm offers a quadratic improvement in the dependence on $1/p$ over classical random sampling, which needs $\Theta(1/p)$ samples; because each quantum query is costly, however, it is faster in total running time only when $p$ is very small (roughly below $N^{-4}$), where the number of queries is no longer constant. We describe the problem formulation, detail the oracle construction (edge-length loading, sequential quantum addition, and threshold comparison), and analyse the query complexity to achieve the required result.
\end{abstract}

\section{Introduction}

In the world of combinatorics, few problems are as clearly applicable in daily life as the travelling salesman problem. From GPS navigation to delivery routes, the quintessential salesman is representative of an array of different problems, each of which traces a Hamiltonian cycle through a series of cities, aiming to keep the distance covered as small as possible. The problem can be posed in several different forms, such as the optimisation version, which aims to find the shortest possible route, or the decision version, which checks whether the tour length is at most equal to a given value. These problems have been tackled through a variety of approaches, including classical algorithms and heuristics. However, with the rise of quantum computing, quantum algorithms may offer new ways to address such complex problems more efficiently. This paper aims to frame the travelling salesman problem as a robust promise problem.

Rather than asking whether a single optimal tour exists, the robust formulation asks a different question: does a significant fraction of all possible tours meet a given distance threshold? The motivation is practical: disruptions such as traffic or weather can make the mathematically optimal route impractical, and if many tours meet the threshold, good alternatives are likely to remain. To apply a quantum algorithm to this question, this paper frames it as a promise problem. Instead of asking about every possible fraction of feasible tours, the problem restricts our attention to two clear cases: either a substantial fraction of tours (at least some fixed proportion $p$) are feasible, or none are, while cases in between are excluded by assumption. This restriction is what allows Grover's algorithm, whose required number of search iterations depends on the fraction of solutions, to solve the problem with few queries. Under this promise, this paper shows that the number of Grover iterations required depends only on $p$, and not on the number of cities involved. The remainder of this paper describes existing approaches to the TSP, mathematically formulates both the standard and robust versions of the problem, describes the oracle construction, derives the resulting query complexity and discusses the results of the study.

\section{Literature review}

Quantum computing has been explored as a way to reduce the time complexity of solving complex problems. Quantum algorithms use properties such as superposition, entanglement and interference, and for some problems they are known to outperform classical algorithms \citep{montanaro2016}, a speedup often referred to as quantum advantage. One of the problems frequently used to explore this potential speedup is the Travelling Salesman Problem.

Studied by mathematicians since at least the 1930s, and famously solved for a 49-city instance by Dantzig, Fulkerson, and Johnson in 1954 \citep{dantzig1954}, the TSP asks the following question: given a set of cities connected by paths of varying lengths, what is the shortest route that visits each city exactly once and returns to the origin? Each possible sequence of cities is called a tour, and each tour corresponds to a Hamiltonian cycle (beginning and ending at the origin), with cities as vertices and paths as edges in the corresponding graph. The decision version of the TSP is NP-complete \citep{garey1979}, and all known exact algorithms have running times that grow exponentially with the number of cities. Given its many applications in operations research, solving the TSP efficiently has remained an active area of research for decades.

Classical approaches to the TSP range from exact algorithms to heuristics. The Held-Karp dynamic programming algorithm, developed in 1962 \citep{heldkarp1962}, achieved a runtime of $O(2^N N^2)$. Alongside exact methods, heuristic methods, including simulated annealing, genetic algorithms, and the Lin-Kernighan method, have been used to arrive at sufficiently accurate solutions, though without guarantees of optimality. These methods can handle larger instances but at the cost of certainty.

As classical algorithms approached their limits, quantum computing emerged as a promising alternative. Grover's algorithm \citep{grover1997} provides a quadratic speedup for unstructured search: it finds a marked item among $T$ items with $O(\sqrt{T})$ queries, compared with $O(T)$ classically, and this has inspired multiple quantum approaches to the TSP. Grover Adaptive Search (GAS) directly applies Grover's algorithm to tours encoded as quantum bit strings; \citet{zhu2022} designed a Hamiltonian Cycle Detection oracle that automatically removes invalid tours, paired with an anchor register strategy that saves qubits, so that a 7-city TSP needs only 31 qubits in their simulation. \citet{ruan2020} applied a hybrid quantum-classical approach, the Quantum Approximate Optimisation Algorithm (QAOA; \citealp{farhi2014}), to the TSP, encoding the constraints of a valid tour in the mixing Hamiltonian and mapping the edges between cities to qubits, which reduces the number of qubits needed. Most recently, \citet{sato2025} introduced Two-Step Quantum Search (TSQS), applying Grover's algorithm in two successive stages, first to prepare a superposition of all valid tours, then to identify the optimal one, using an encoding known as higher-order unconstrained binary optimisation (HOBO) to reduce the number of qubits required.

A different approach was taken by \citet{ambainis2019}, who sped up the Held-Karp dynamic programming algorithm itself instead of searching over tours directly. By using Grover's algorithm within the dynamic programming steps, they reported a runtime of $O^*(1.728^N)$ for the standard TSP, with no promise or special structure required. However, this method assumes quantum random-access memory (QRAM), a quantum memory that can be accessed in superposition, which is not yet practical to build.

Despite these innovations, the quantum approaches described above have so far been demonstrated only on small instances, such as the seven-city example of \citet{zhu2022}, far below the sizes that classical heuristics handle. This is largely due to the limitations of current quantum hardware. \citet{preskill2018} describes the present period as the Noisy Intermediate-Scale Quantum (NISQ) era, in which devices have limited numbers of qubits and no error correction, so that noise limits the size of the circuits that can be run reliably.

This raises the question of whether Grover's algorithm, while limited in accelerating the standard TSP, might offer greater advantage when applied to a different class of problem. Robust optimisation, which seeks solutions that perform reliably across a range of uncertain conditions rather than a single optimal solution under fixed assumptions, has been effectively applied in classical settings such as traffic signal control and vehicle routing \citep{yin2008,gounaris2013}. The variant studied in this paper is inspired by this idea rather than being a robust optimisation problem in the usual sense: instead of protecting a single tour against uncertain edge lengths, it asks whether a large fraction of all tours meet the threshold, so that good alternatives are likely to remain if some routes become impractical. This reformulation changes the task given to Grover's algorithm: instead of finding the single best tour, it only has to find one of many feasible tours. Whether this gives an advantage over classical methods is examined in Section \ref{sec:discussion}.

\section{Problem Formulation}

\subsection{The decision framing of the classical TSP}

A tour can be defined as an ordered sequence of cities, visiting each city exactly once. The aim of the problem is to minimise the total distance covered in the tour. The Travelling Salesman Problem is represented as a graph problem. Consider a graph $G = (V, E)$, where $V$ represents the set of $N$ cities in the tour and $E$ represents all edges connecting pairs of cities. Each edge has a known associated distance $d_{ij}$ where $(i,j) \in E$. Here, $\pi$ represents a tour permutation of $N$ cities. The optimisation problem may be represented as follows:

\begin{equation}
\min \sum_{i=0}^{N-1} d_{\pi(i),\pi(i+1)}
\end{equation}

where $\pi(N) = \pi(0)$ as the salesman returns to the original node upon completing the tour.

This optimisation problem can be reduced to an equivalent decision problem. In this case, the decision problem involves determining whether there exists a tour of cities that satisfies the following condition:

\begin{equation}
\exists\, \pi \in S_N \;\Big|\; \sum_{i=0}^{N-1} d_{\pi(i),\pi(i+1)} \leq K, \qquad K \in \mathbb{N}^+
\label{eq:decision}
\end{equation}

Where $\pi(N) = \pi(0)$, $S_N$ represents the set of all possible permutations of the $N$ cities of the tour, and $K$ is the maximum threshold value for the path length covered in the tour.

\subsection{The Robust TSP as a promise problem}

The robust TSP chooses to ask, not whether a single feasible tour exists, but if the fraction of feasible tours is significant. In real life, this means that the decision still holds true even if there are some interruptions or perturbations that impact the tours, rendering them impractical. Writing $L(\pi)$ for the total length of tour $\pi$ (the sum in Equation \ref{eq:decision}), let $M(K)$ denote the number of feasible tours:

\begin{equation}
M(K) = \big|\{\, \pi \in S_N \;:\; L(\pi) \leq K \,\}\big|
\end{equation}

The definition of the problem establishes a robustness parameter $p \in (0,1]$. Here $p$ is the minimum fraction of tours that must be feasible: we ask whether at least a fraction $p$ (that is, $100p\%$) of all possible tours are feasible. This fraction, a constant, may be decided in advance and is independent of the number of cities.

As a promise problem, the input comes with the guarantee that either $M(K) \geq p \cdot N!$ or $M(K) = 0$. The algorithm must decide which case holds. Instances with $0 < M(K) < p \cdot N!$ are excluded by the promise. If the promise is violated, the algorithm still never accepts an infeasible tour, because every output is checked classically to confirm that it is a valid tour of length at most $K$, but it may answer ``no'' even though some feasible tours exist.

\subsection{Feasibility Function}

Over the set of tours $S_N$ we construct a feasibility function $\kappa$ based on Equation \ref{eq:decision} such that

\begin{equation}
\kappa(\pi) =
\begin{cases}
1 & \text{if } L(\pi) \leq K \\
0 & \text{otherwise}
\end{cases}
\end{equation}

This will be implemented in a quantum routine to solve this definition of the travelling salesman problem.

\section{Quantum Algorithm for Robust Travelling Salesman Problem}
\label{sec:algorithm}

The quantum routine for the robust Travelling Salesman Problem utilises Grover's algorithm to search for tours whose total path length is at most a given threshold $K$. The overall approach involves implementing the oracle, which evaluates the feasibility function $\kappa$ for all tours at once, and then implementing Grover's search, which increases the probability of measuring a feasible tour. The registers used by the circuit are described first, following which the algorithm is outlined step by step.

\subsection{Registers and notation}

The circuit can be described in terms of five groups of qubits, referred to as registers. Consider first the $N$ node registers, each consisting of $\lceil \log_2 N \rceil$ qubits, where node register $i$ represents the city $\pi(i)$ visited at position $i$ of the tour. Together, the node registers are written as $|\pi\rangle = |\pi(0), \pi(1), \ldots, \pi(N-1)\rangle$. The $N-1$ offset registers $c(0), \ldots, c(N-2)$ are used in preparing the tours in step 1. Further, there are $N$ distance registers $d_0, \ldots, d_{N-1}$ of $w$ qubits each, where $w$ represents the number of bits needed to write the largest edge length. Here, register $d_i$ stores the length of the $i$-th edge of the tour, denoted $\ell_i(\pi) = d_{\pi(i),\pi(i+1)}$, and the distance registers together are written as $|\ell(\pi)\rangle = |\ell_0(\pi), \ldots, \ell_{N-1}(\pi)\rangle$. It may be noted that a register $d_i$ carries one index, whereas a distance $d_{uv}$ carries two. Finally, the sum register $s$ stores the running total of the edge lengths, and the feasibility qubit $f$ stores the value of $\kappa(\pi)$.

The states in Equations \ref{eq:prep}--\ref{eq:feas} are always written in the same order, namely the node registers, the distance registers, the sum register and the feasibility qubit, with a subscript ($n$, $d$, $s$ or $f$) indicating the register in question. Each sum runs over all $N!$ tours $\pi \in S_N$. For readability, the offset registers are not shown; since each tour is paired with exactly one setting of the offsets, leaving them out does not change any of the amplitudes.

\subsection{Steps of the algorithm}

\begin{figure}[htbp]
\centering
\includegraphics[width=0.95\textwidth]{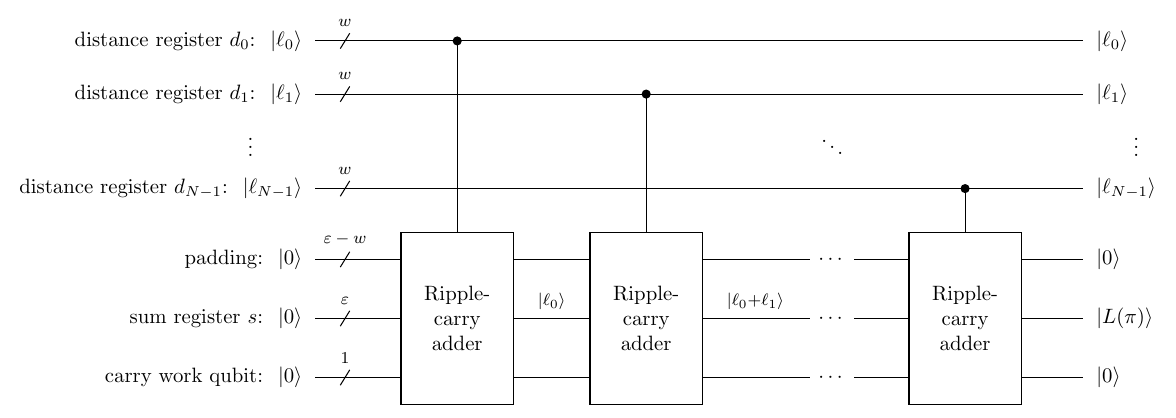}
\caption{Quantum circuit for sequential addition (step 3). Each ripple-carry adder adds one distance register into the sum register. A dot joined to a box marks a register that the box reads but does not change. The label above each slash gives the number of qubits on that wire.}
\label{fig:adder}
\end{figure}

\begin{figure}[htbp]
\centering
\includegraphics[width=\textwidth]{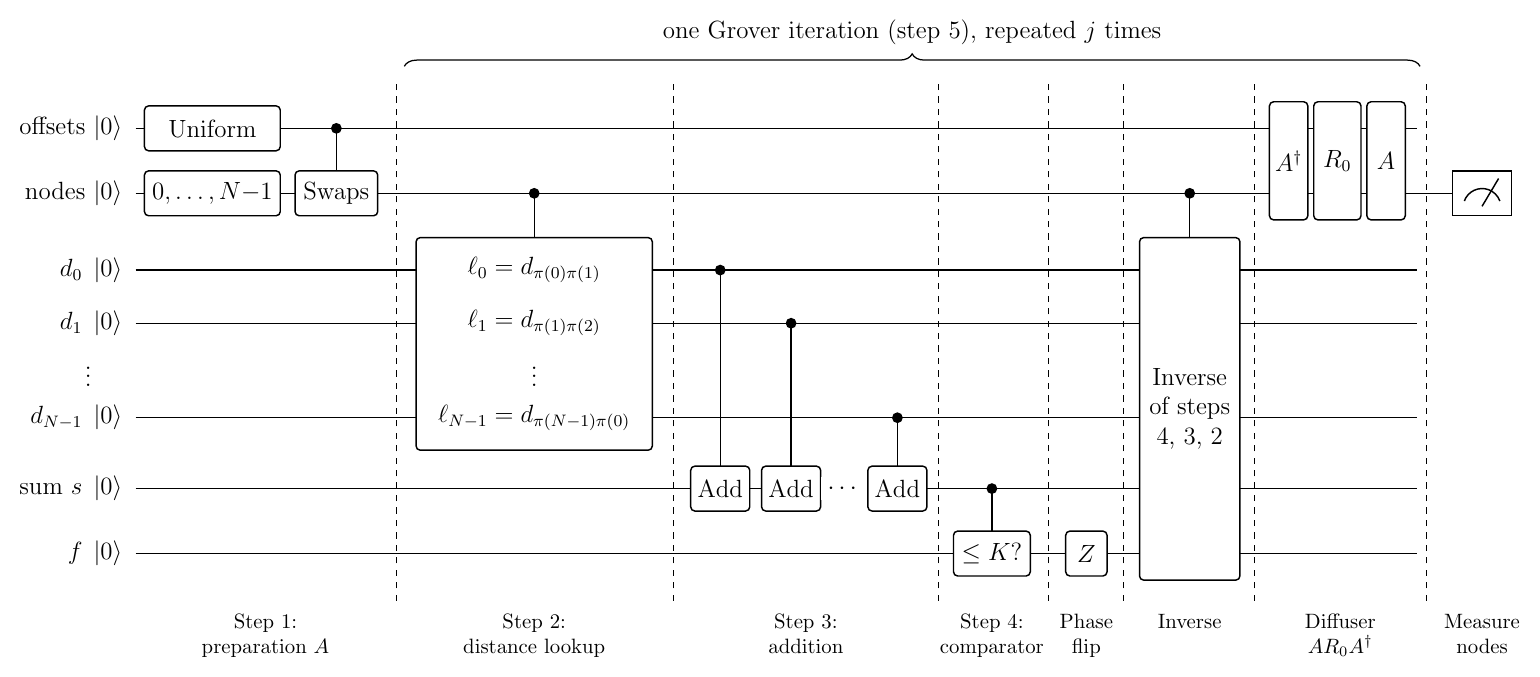}
\caption{Quantum circuit for the robust Travelling Salesman Problem. A dot joined to a box marks a register that the box reads but does not change. In step 1, `Uniform' puts each offset register into an equal superposition of its allowed values, `$0, \ldots, N-1$' writes the starting order into the node registers, and `Swaps' performs the controlled swaps; together these form the preparation circuit $A$. In the diffuser, $R_0 = 2|0\rangle\langle 0| - I$. The padding and work qubits of Figure \ref{fig:adder} are not shown.}
\label{fig:circuit}
\end{figure}

\begin{enumerate}
\item \textbf{Preparing the tours.} All qubits begin in the state 0, and the fixed order $(0, 1, \ldots, N-1)$ is first written into the node registers using $X$ gates. Each offset register $c(i)$, which can take the values $0, 1, \ldots, N-1-i$, is then put into an equal superposition of its allowed values; because $N-i$ is usually not a power of two, this uses a standard rotation-based circuit rather than plain Hadamard gates. Following this, for $i = 0, \ldots, N-2$, node register $i$ is swapped with node register $i + c(i)$, controlled on the value of $c(i)$, which acts as a quantum version of the Fisher--Yates shuffle. Each choice of offsets gives exactly one permutation, so this preparation circuit, which we call $A$, produces an equal superposition of all $N!$ valid tours, each paired with its offsets, using $O(N^2 \log N)$ gates. The resulting state may be represented as follows:

\begin{equation}
\frac{1}{\sqrt{N!}} \sum_{\pi \in S_N} |\pi\rangle_n \, |0\rangle_d \, |0\rangle_s \, |0\rangle_f
\label{eq:prep}
\end{equation}

\item \textbf{Loading the edge lengths.} Apply distance-lookup oracles for each of the edges $(\pi(i), \pi(i+1))$ of the tour, where position $N$ refers back to position $0$ as the salesman returns to the origin. For each possible pair of cities $(u,v)$, a gate controlled on $\pi(i) = u$ and $\pi(i+1) = v$ writes the binary value of the distance $d_{uv}$ into the distance register $d_i$. If the graph is not complete, pairs that are not joined by an edge are given a length larger than $K$, so that tours using them are never marked feasible. This is done individually for each edge, and all the distance outputs are stored in their individual registers:

\begin{equation}
\frac{1}{\sqrt{N!}} \sum_{\pi \in S_N} |\pi\rangle_n \, |\ell(\pi)\rangle_d \, |0\rangle_s \, |0\rangle_f
\label{eq:load}
\end{equation}

\item \textbf{Adding the edge lengths.} This routine adds the distances associated with the edges sequentially to find the total path length, which is stored in the sum register. Quantum adders work on binary numbers, so the distances and the running sum are stored in binary. Each addition is carried out by a ripple-carry adder \citep{cuccaro2004}, which adds the number held in one register to the number held in a second register while leaving the first unchanged, $|a\rangle|b\rangle \rightarrow |a\rangle|a+b\rangle$. Similar to column addition by hand, the adder begins at the least significant bit and passes a carry from each bit to the next, for which one additional work qubit is required.

The size of the sum register can be found beforehand by noting the number of edges in a tour and the largest edge length $d_{\max}$. As the total length is at most $N \cdot d_{\max}$, a register of $\varepsilon = \lfloor \log_2(N \cdot d_{\max}) \rfloor + 1$ qubits is sufficient, and the sum never overflows. Since a distance register consists of only $w$ qubits, $\varepsilon - w$ padding qubits in the state 0 are appended in front of it so that the total number of qubits matches that of the sum register. The padding qubits are never changed and the work qubit is returned to 0 after each addition, allowing the same qubits to be reused for the next edge. For the first edge, the summation with 0 leads to the distance of the first edge:

\begin{equation}
\frac{1}{\sqrt{N!}} \sum_{\pi \in S_N} |\pi\rangle_n \, |\ell(\pi)\rangle_d \, |\ell_0(\pi)\rangle_s \, |0\rangle_f
\label{eq:firstadd}
\end{equation}

To add the lengths of all edges, the above step is repeated for each of the $N$ edges of the tour, giving the total length $L(\pi) = \sum_{i=0}^{N-1} \ell_i(\pi)$:

\begin{equation}
\frac{1}{\sqrt{N!}} \sum_{\pi \in S_N} |\pi\rangle_n \, |\ell(\pi)\rangle_d \, |L(\pi)\rangle_s \, |0\rangle_f
\label{eq:sum}
\end{equation}

An example circuit to demonstrate this step is given in Figure \ref{fig:adder}.

\item \textbf{Feasibility function.} To determine the feasible tours such that the total distance is at most the threshold $K$, an integer comparator is applied, comparing the value in the sum register to the given threshold value. This sets the feasibility qubit $f$ to 1 if the threshold condition $L(\pi) \leq K$ is satisfied and leaves it at 0 otherwise, so that $f$ stores $\kappa(\pi)$:

\begin{equation}
\frac{1}{\sqrt{N!}} \sum_{\pi \in S_N} |\pi\rangle_n \, |\ell(\pi)\rangle_d \, |L(\pi)\rangle_s \, |\kappa(\pi)\rangle_f
\label{eq:feas}
\end{equation}

\item \textbf{Grover's search.} Grover's algorithm is applied to find the tours satisfying the threshold condition. This requires a phase flip on the feasibility qubit, implemented by a $Z$ gate, which multiplies every feasible tour by $-1$ in order to `mark' the good states while leaving the other tours unchanged. The inverse of steps 4, 3 and 2 is then applied to return the feasibility, sum and distance registers back to 0, with the marked tours retaining their minus sign. A Grover diffuser, $A R_0 A^\dagger$, is then applied to the offset and node registers, where $R_0 = 2|0\rangle\langle 0| - I$ leaves the all-zero state unchanged and flips the sign of every other state. The diffuser reflects the state about the superposition prepared in step 1, amplifying the marked good states. Steps 2--4, the phase flip, the inverse and the diffuser together constitute one Grover iteration, as shown in Figure \ref{fig:circuit}. The number of iterations is chosen at random, and each run also checks one random tour classically, as explained in Section \ref{sec:query}. The node registers are then measured, and the result is checked classically: it must be a valid tour with total length at most $K$.
\end{enumerate}

\section{Complexity Analysis}

\subsection{Query Complexity}
\label{sec:query}

According to \citet{boyer1998}, Grover's search algorithm searching a space of total $T$ states with $M$ marked (favourable) states can be described in terms of an angle $\theta$, defined by $\sin^2\theta = M/T$. Before any Grover iteration is applied, a measurement returns a marked state with probability $\sin^2\theta = M/T$, which is no better than a random guess. Each Grover iteration rotates the state by an angle $2\theta$ towards the marked states, so that after $j$ iterations the probability of measuring a marked state may be represented as follows:

\begin{equation}
P(j) = \sin^2\big((2j+1)\theta\big)
\label{eq:prob}
\end{equation}

This probability is highest when $(2j+1)\theta$ is close to $\pi/2$, which leads to the number of iterations, $r$, used by \citet{boyer1998}:

\begin{equation}
r = \left\lfloor \frac{\pi}{4\theta} \right\rfloor
\label{eq:iters}
\end{equation}

In the robust TSP, the search space is the set of all valid tours, so $T$ equals $N!$. Further, according to the promise defined previously, the number of feasible tours satisfies the relation $M(K) \geq p \cdot N!$. Here, $\sin\theta = \sqrt{M/T}$, and since $\theta \geq \sin\theta$ for any angle $\theta \geq 0$, substituting into Equation \ref{eq:iters} gives:

\begin{equation}
r \leq \frac{\pi}{4\theta} \leq \frac{\pi}{4} \sqrt{\frac{T}{M}} \leq \frac{\pi}{4} \sqrt{\frac{N!}{p \cdot N!}} = \frac{\pi}{4} \sqrt{\frac{1}{p}}
\label{eq:bound}
\end{equation}

where the factor $N!$ cancels. From this, we infer that

\begin{equation}
r \leq \frac{\pi}{4} \cdot \frac{1}{\sqrt{p}} \;\Rightarrow\; r = O\!\left(\frac{1}{\sqrt{p}}\right)
\end{equation}

If $M(K)$ is known exactly, $r$ iterations would be enough to find a feasible tour with probability at least $1/2$, with this probability approaching 1 when the feasible tours form only a small fraction of all tours. The number of iterations is independent of the total number of cities $N$, and is only defined by $p$, the promised robustness level whose value is a pre-decided constant. For example, if $p = 0.05$, then $r \leq 3.51$, meaning that at most 3 iterations are needed regardless of whether the instance involves 5 cities or 5{,}000.

In practice, however, $M(K)$ is not known exactly, as the promise only guarantees that it is at least $p \cdot N!$. This matters because the probability in Equation \ref{eq:prob} rises and then falls again as $j$ increases. If the true fraction of feasible tours is larger than $p$, then $\theta$ is larger than expected, and a number of iterations chosen for the fraction $p$ can rotate the state past the feasible tours, causing the success probability to fall close to zero even though the promise holds. To avoid this, this paper follows \citet[Lemma 2]{boyer1998}. For $p \geq 1/2$, no quantum search is required, since a single random tour is already feasible with probability at least $1/2$. For $p < 1/2$, each run of the algorithm involves two actions. First, one tour is picked at random and checked classically. Second, a number $j$ is chosen uniformly at random from $\{0, 1, \ldots, m-1\}$, where $m = \left\lceil 1/(2\sqrt{p(1-p)}) \right\rceil$, following which $j$ Grover iterations are performed, the node registers are measured and the result is checked classically. The run is considered successful if either check finds a feasible tour.

Lemma 2 of \citet{boyer1998} states that, averaged over the random choice of $j$, the second action succeeds with probability at least $1/4$ whenever $m \geq 1/\sin(2\theta)$. Let $t = M(K)/N!$ represent the true fraction of feasible tours, so that $\sin(2\theta) = 2\sin\theta\cos\theta = 2\sqrt{t(1-t)}$. In this case, the condition $m \geq 1/\sin(2\theta)$ holds whenever $t(1-t) \geq p(1-p)$, that is, whenever $p \leq t \leq 1-p$. The only remaining case is $t > 1-p$, in which more than half of all tours are feasible, so the classical check in the first action succeeds with probability greater than $1/2$. Therefore, in every case, a run succeeds with probability at least $1/4$ on any instance satisfying the promise.

Each run uses fewer than $m$ Grover iterations, where $m \leq \lceil 1/\sqrt{2p}\, \rceil$ because $1-p > 1/2$. For the robust promise problem, Grover's query complexity is therefore $O(1/\sqrt{p}) = O(1)$ for a fixed value of $p$. To increase the accuracy of the result, the process can be repeated. Since a single run fails with probability at most $3/4$, $k$ independent runs all fail with probability at most $(3/4)^k$, and choosing $k \geq \log(1/\delta)/\log(4/3)$, which is $O(\log(1/\delta))$, reduces this to at most $\delta$, where $\delta$ is the acceptable probability of failure chosen in advance. The outcome is accepted if any run returns a tour that can be verified for feasibility, which boosts the overall success probability to $1 - \delta$. It may also be noted that if no feasible tour exists, no run can return one, so the answer ``no'' is always correct in that case.

\subsection{Gate and Qubit Costs}
\label{sec:cost}

The dominant cost in each Grover iteration is loading the edge lengths (step 2 of Section \ref{sec:algorithm}). For each of the $N^2$ possible city pairs $(u,v)$, a gate controlled on $\pi(i) = u$ and $\pi(i+1) = v$ writes the $w$-bit value of $d_{uv}$ into the distance register; this costs $O(N^2(\log N + w))$ gates per edge, or $O(N^3(\log N + w))$ for all $N$ edges. By comparison, adding the $N$ edge lengths with ripple-carry adders costs only $O(N(w + \log N))$ gates, because each of the $N$ additions works on $\varepsilon = O(w + \log N)$ bits and a ripple-carry adder uses a fixed number of gates per bit. The preparation circuit $A$ and its inverse (used in the diffuser) cost $O(N^2 \log N)$. Including uncomputation, one Grover iteration therefore uses $O(N^3(\log N + w))$ gates, which is $O(N^4)$ whenever $w \leq N$. In terms of qubits, the algorithm requires $N$ registers of $\lceil \log_2 N \rceil$ qubits each to encode the tour permutation, $N-1$ offset registers of at most $\lceil \log_2 N \rceil$ qubits each, $N$ registers of $w$ qubits to store the individual edge distances, a sum register of size $\varepsilon = \lfloor \log_2(N \cdot d_{\max}) \rfloor + 1$ qubits, where $d_{\max}$ is the largest edge length, to store the total tour length, $\varepsilon - w$ padding qubits and one work qubit for the adders, and a single feasibility qubit. The total qubit count is therefore $O(N(\log N + w))$, close to linear in $N$.

Combining this with the query complexity derived in Section \ref{sec:query}, each run of the algorithm uses $O(N^3(\log N + w)/\sqrt{p})$ gates, which is $O(N^4/\sqrt{p})$ when $w \leq N$: it performs $O(1/\sqrt{p})$ Grover iterations, each costing $O(N^3(\log N + w))$ gates, mainly for loading the edge lengths. Repeating the run $O(\log(1/\delta))$ times multiplies this by a factor of $\log(1/\delta)$.

\section{Discussion and Limitations}
\label{sec:discussion}

The central result of this paper is that, under the promise defined earlier, Grover's algorithm can be used as an approach to the robust TSP using a number of oracle queries that depends only on the robustness parameter $p$, not on the number of cities $N$. Reframing the problem around how many tours are feasible, rather than finding a single best one, enables us to use Grover's algorithm with an oracle built from simple components to tackle the problem.

The following two points clarify what this result means. First, the constant-query guarantee applies to the number of Grover iterations, not the total running time. Each query costs at most $O(N^4)$ gates due mainly to loading the edge lengths, so the overall gate complexity is $O(N^4/\sqrt{p})$, which is polynomial in $N$. The number of iterations does not grow with $N$, but as discussed below, this alone does not make the algorithm faster than classical methods. Second, the guarantee only holds for instances that satisfy the promise that either a large enough fraction of tours (at least $p$) is feasible, or none are at all. This is a deliberate simplification, and a similar promise-based approach is used in other quantum algorithms, such as the Deutsch--Jozsa algorithm \citep{deutsch1992}, to obtain clean complexity results.

It is important to compare this with a classical algorithm under the same promise. The simplest classical method involves picking tours at random and checking each one. If at least a fraction $p$ of the tours is feasible, each random tour is feasible with probability at least $p$, so the probability that $k$ random tours are all infeasible is at most $(1-p)^k$. Taking $k$ to be about $1/p$ already brings this below $1/e \approx 0.37$, and a few times more makes it as small as required. With far fewer than $1/p$ samples, on the other hand, the search will usually miss every feasible tour when only a fraction $p$ of them is feasible, and the same holds for any classical method that learns about the instance only by checking tours one at a time. Classical sampling therefore requires about $1/p$ checks, whereas Grover's algorithm requires about $1/\sqrt{p}$ iterations, which in terms of queries is a quadratic improvement. This is the general speedup Grover's algorithm offers for any search problem, rather than an advantage specific to the TSP.

However, for a fixed $p$, this improvement does not carry over to running time, as a quantum query is far more costly than a classical check. Checking one tour classically involves looking up $N$ edge lengths and adding them, which takes $O(N(\log N + w))$ bit operations, whereas one Grover iteration costs $O(N^3(\log N + w))$ gates (Section \ref{sec:cost}), roughly $N^2$ times more. The total cost of classical sampling is therefore about $\frac{1}{p} \cdot N(\log N + w)$, while that of Grover's algorithm is about $\frac{1}{\sqrt{p}} \cdot N^3(\log N + w)$. In this case, the quantum cost is the smaller of the two only when $1/\sqrt{p}$ exceeds roughly $N^2$, that is, when $p$ is below about $N^{-4}$. For $N = 10$ cities, this means that fewer than one tour in $10{,}000$ is feasible, in which case about $1/\sqrt{p} = 100$ or more Grover iterations are needed. A value of $p$ small enough to give a speedup must therefore shrink as $N$ grows, rendering the number of Grover iterations no longer constant. For this circuit, compared with classical random sampling, a constant number of queries and a genuine speedup therefore never hold at the same time.

The extra factor of $N^2$ arises from loading the edge lengths with one controlled gate for every pair of cities. If quantum random-access memory were available, the circuit could look up each edge length directly, and each Grover iteration would cost about as much as one classical check, allowing the quadratic improvement in $1/p$ to carry over to running time. For a fixed $p$, however, this is still only a constant factor, and quantum random-access memory is a strong hardware assumption.

Further, random sampling is only the simplest classical baseline, as practical classical heuristics can exploit the structure of the distances directly. As with the Deutsch--Jozsa problem, which a classical randomised algorithm can also solve efficiently if a small probability of error is allowed, it is the promise that makes the problem easy. The promise is also demanding in practice: for thresholds $K$ close to the optimal tour length, the fraction of feasible tours is typically tiny and shrinks as $N$ grows, so a constant $p$ corresponds to thresholds close to the length of a typical tour.

The promise used here, that either at least a fraction $p$ of candidate tours are feasible or none are, is what allows simple random sampling to succeed with about $1/p$ samples, with amplitude amplification, the general form of Grover's algorithm, reducing this to about $1/\sqrt{p}$ queries \citep{brassard2002}. To our knowledge, the contribution of this paper is to apply this promise to a threshold version of the TSP, motivated by robustness to disruptions, and to give an explicit circuit construction and cost analysis for it.

This paper focuses on deriving the algorithm and its query complexity theoretically. Future steps would be to verify the result through simulation and implementation on actual quantum hardware.

\section{Conclusion}

The travelling salesman problem has long stood as a benchmark for testing the limits of computation, and more recently, the promise of quantum algorithms. By taking a different route with the framing of the question as a robust promise problem, the paper was able to apply Grover's algorithm to a version of the TSP in which feasible tours are plentiful, where it needs far fewer queries than it does for the standard TSP. Rather than searching through nearly every possible tour, it was demonstrated that Grover's algorithm can approach the robust TSP using a number of queries that depends only on the robustness parameter $p$, and not on the number of cities involved. This shows that a small shift in how a problem is framed, from finding the single best answer to asking whether good answers are plentiful, can greatly reduce the number of queries a search needs, although, as Section \ref{sec:discussion} shows, the same shift also makes the problem easy for classical random sampling. The next step would be to verify this result on a simulator and, eventually, on real quantum hardware. Similar approaches could be explored for other combinatorial problems, provided that each is compared fairly with classical methods such as random sampling.

\section*{Acknowledgements}

I would like to thank my mentor, Professor Vinayak Dixit at the University of New South Wales, under whose guidance this entire project was carried out. I would also like to thank my family and my teachers at The Shri Ram School Aravali for supporting my journey. Claude, an AI assistant developed by Anthropic, was used to explain concepts, verify the mathematics, circuit figures and references, and help with the wording and language of this paper. I take responsibility for the content of the paper.

\bibliographystyle{plainnat}

\end{document}